\documentclass{PoS}

\usepackage{amsmath}
\usepackage{amssymb}
\usepackage{graphicx}

\newcommand{\ii}{\ensuremath{\textrm{i}}}

\leftline{\parbox{11cm}{\large\rm BI-TP-2010/28, HU-EP-10/50}}
\vspace*{-0.5cm}

\title{Thermal transition temperature from twisted mass QCD}

\ShortTitle{Thermal transition temperature from twisted mass QCD}

\author{tmfT collaboration:}
\author{Florian Burger, Malik Kirchner, Michael M\"uller-Preussker\\
        Humboldt-Universit\"at zu Berlin, Institut f\"ur Physik, 12489  Berlin, Germany\\
        E-mail: \email{burger@physik.hu-berlin.de}\\
E-mail: \email{malik@physik.hu-berlin.de}\\
E-mail: \email{mmp@physik.hu-berlin.de}
}

\author{Ernst-Michael Ilgenfritz\\
        Universit\"at Bielefeld, Fakult\"at f\"ur Physik, 33615 Bielefeld, Germany \\
        Humboldt-Universit\"at zu Berlin, Institut f\"ur Physik, 12489  Berlin, Germany\\
        E-mail: \email{ilgenfri@physik.hu-berlin.de}}

\author{Maria Paola Lombardo\\
        Laboratori Nazionali di Frascati, INFN, 100044 Frascati, Roma, Italy\\
        E-mail: \email{mariapaola.lombardo@lnf.infn.it}}

\author{Owe Philipsen, \speaker{Lars Zeidlewicz}\\
        Goethe-Universit\"at Frankfurt, Insitut f\"ur Theoretische Physik, 60438 Frankfurt am Main, Germany\\
        E-mail: \email{philipsen@th.physik.uni-frankfurt.de}\\
E-mail: \email{zeidlewicz@th.physik.uni-frankfurt.de}
}

\author{Carsten Urbach\\
        Universit\"at Bonn, HISKP and Bethe Center for Theoretical Physics, 53115 Bonn, Germany\\
        E-mail: \email{urbach@hiskp.uni-bonn.de}}

\abstract{We present the current status of lattice simulations with $N_f=2$ maximally twisted mass Wilson fermions at finite temperature. In particular, the determination of the thermal transition temperature is discussed.}

\FullConference{The XXVIII International Symposium on Lattice Filed Theory\\
                 June 14-19,2010\\
                 Villasimius, Sardinia Italy}

\graphicspath{{./figures/}}

\begin{document}

\section{Introduction}
One of the main aspects to be studied in thermal quantum chromodynamics (QCD) is the transition -- or smooth crossover -- from a chirally symmetric and deconfined region of phase space to a hadronic phase with broken chiral symmetry. 
This transition has occurred in the cooling of the early universe and is hoped to be reproduced in heavy ion collision experiments. 

\begin{figure}[h]
\centering
\includegraphics[width=.4\textwidth]{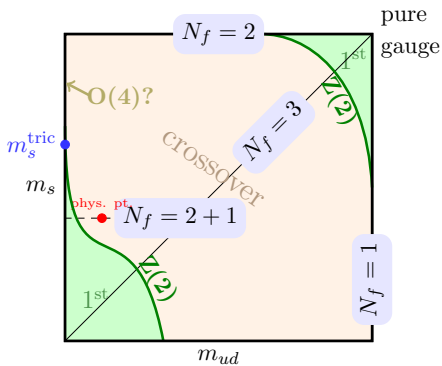}
\caption{Nature of the thermal transition in the quark mass plane.}
\label{fig:columbia}
\end{figure}

A lot of effort has been put into lattice studies of this transition, see e.\,g.~\cite{DeTar:2008qi} and the most recent overview by K.\,Kanaya~\cite{kanaya:lat2010}. Hybrid Monte-Carlo simulations are restricted to vanishing or at most small chemical potential due to the sign problem that otherwise prohibits importance sampling. 
In case of zero chemical potential and physical quark masses, it has been demonstrated  that the thermal transition is no true phase transition but really a smooth crossover~\cite{Aoki:2006we}. 
The current knowledge about the nature of the phase transition for different quark masses is shown in Figure~\ref{fig:columbia}. 
This dependence is particularly important for lattice simulations with Wilson-type fermions, for which physical quark masses have not been feasible so far and extrapolations are necessary.
Furthermore, by knowing the qualitative properties of this plot, some constraints on the enlarged phase diagram including a non-vanishing chemical potential can be anticipated.

The critical temperature for a true phase transition is uniquely determined by non-analyticities in the partition function. For a crossover there are only pseudo-critical temperatures which depend on the observable. Here we use the Polyakov loop and its susceptibility and compare to the pion norm and the screening mass in the pseudo-scalar channel.

The aforementioned lattice results have been obtained from simulations in which predominantly staggered fermions have been applied. However, the question to what extent this type of fermions is reliable has still not been completely settled.  The main problem is the so-called rooting procedure, i.\,e. the need to take the fourth root of the staggered fermion determinant in order to describe a single flavour of quarks, see e.\,g.~\cite{Creutz:2008nk}.
Therefore, it is highly desirable to use different fermion formulations so that the possible caveats are under better mutual control. 
Wilson-type fermions have very different systematic effects from the staggered formulation. 
In particular, they do not suffer from the so-called taste-breaking that needs to be controlled in staggered simulations, see~\cite{bazavov:lat2010}. 
The major problems of Wilson-type fermions are of course the explicit chiral symmetry breaking and the introduction of $\mathcal{O}(a)$ effects.

In this contribution we report on our ongoing investigation of the thermal transition of QCD applying so-called maximally twisted-mass fermions. These Wilson-type fermions offer automatic $\mathcal{O}(a)$ improvement. The quark mass is determined by the multiplicatively renormalisable twisted mass parameter $\mu$. For an introduction, see~\cite{Shindler:2007vp}.

We use one doublet of degenerate fermions for the twisted mass action. Thus, in terms of Fig.~\ref{fig:columbia}, we are restricted to the upper boundary where the strange quark is infinitely heavy. Here the situation may well be different from that at the physical point.
However, the $N_f=2$ setup is somewhat natural to start from with our action as the twisting always acts on fermion doublets. 
Furthermore, there remains the question whether the thermal transition in the two-flavour chiral limit is really of second order in the $O(4)$ universality class.
Therefore, the chosen setup is very well suited to establish further information about the quark mass dependence of the QCD transition.

Besides our approach relying on the twisted mass formulation, also clover-improved Wilson fermions are currently used for finite temperature studies~\cite{Bornyakov:2009qh,Ejiri:2009hq}, see also~\cite{bornyakov:lat2010,brandt:lat2010} in these proceedings.
As compared to our previous work~\cite{MullerPreussker:2009iu,Ilgenfritz:2009ns}, we now report on simulations with light quark masses $\gtrsim 300$\,MeV that will give information on the quark mass and lattice spacing dependence of the critical temperature.

In the following section, we present the maximally twisted mass lattice formulation as used for our simulations. Furthermore, we describe the non-trivial $(\kappa,\beta,\mu)$ phase diagram and explain our realisation of scans in $\beta$. 
Section~\ref{sec:thermal} gives a summary of our results obtained so far. 
We show the Polyakov loop and the flavour non-singlet pseudo-scalar screening mass as examples for the observed signals and collect the transition temperatures available from our data. This also allows for a discussion of cutoff effects.
Finally, in Section~\ref{sec:concl} we conclude and describe our plans for the continuation of this work.

\section{Lattice action}
For our simulations we rely on zero temperature information by the European Twisted Mass Collaboration (ETMC)~\cite{Baron:2009wt,personal}. The gauge action is chosen to be tree-level Symanzik improved. The maximally twisted fermion matrix takes the form
\begin{equation}
M_\text{mtm} = 1 - \kappa_c D_\text{Wilson} + 2\ii\kappa_c a\mu\gamma_5\tau^3 
\end{equation}
and acts on one fermion doublet. The Pauli matrix $\tau^3$ acts in flavour space and $\kappa_c$ is the hopping parameter at its critical value. For $\mu=0$ the action has its standard Wilson form. In the continuum, the new term $\ii a\mu\gamma_5\tau^3$ can be obtained by a non-anomalous chiral-flavour rotation. However, this rotational symmetry is broken by the Wilson-term so that one has different cutoff effects for each value of this so-called twist-angle. 
For maximal twist, i.\,e. if $\kappa=\kappa_c$, it has been shown that non-vanishing physical observables are automatically $\mathcal{O}(a)$ improved (see~\cite{Shindler:2007vp} and references therein).  We have checked that this remains true at finite temperature by means of a quenched analysis~\cite{MullerPreussker:2009iu}  and a perturbative calculation~\cite{Philipsen:2008gq}.

\begin{figure}
\centering
\includegraphics[width=.4\textwidth,angle=-90]{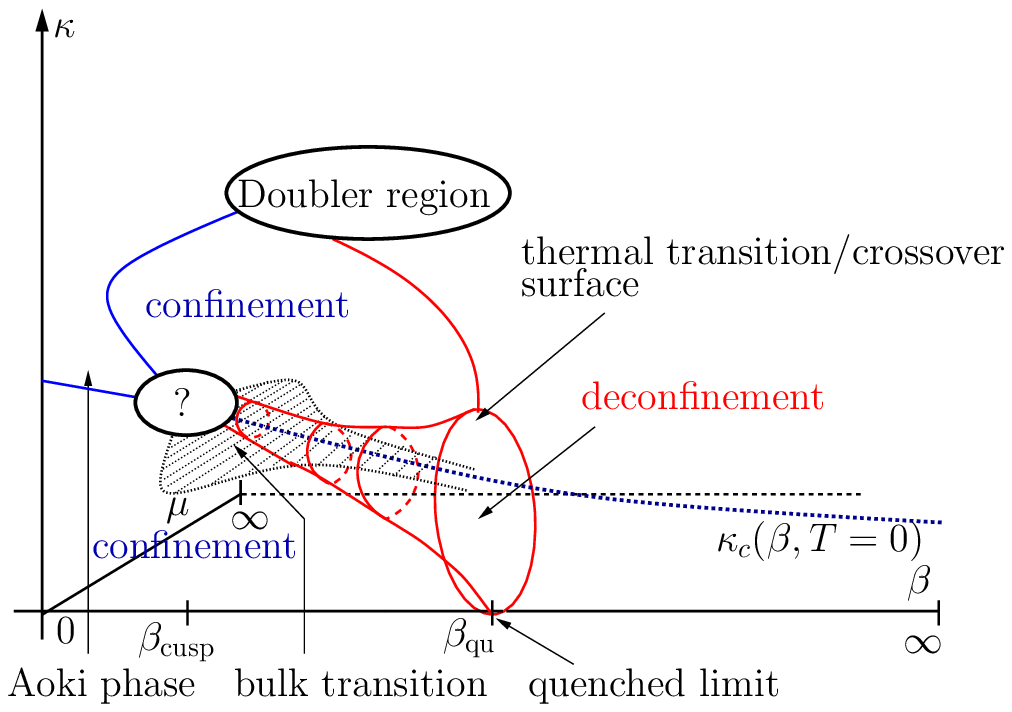}
\hspace{-1.2cm}\includegraphics[width=.35\textwidth,angle=-90]{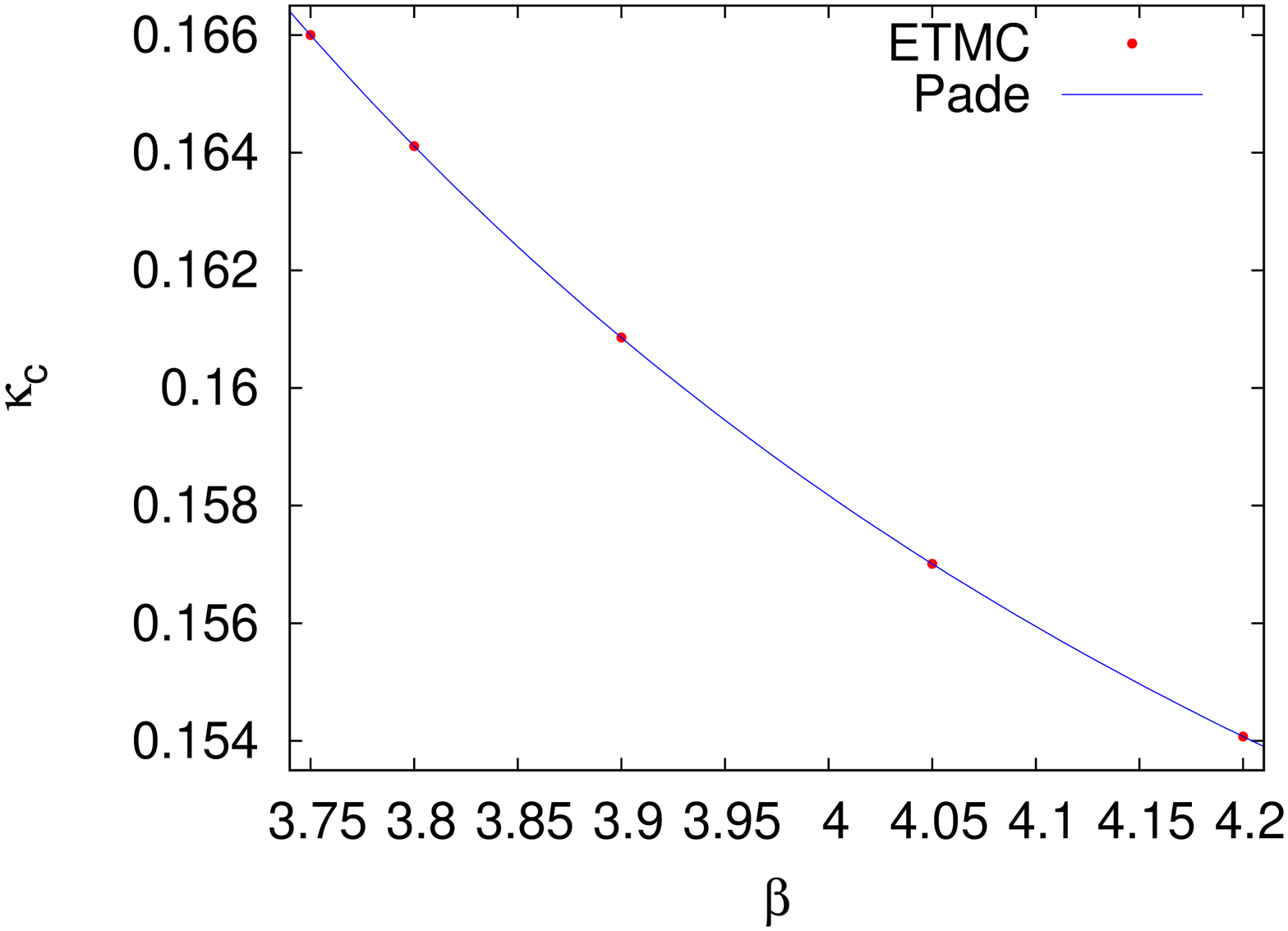}
\caption{Left: Phase diagram in $(\kappa,\beta,\mu)$-space for twisted mass lattice fermions at finite temperature. Right: Interpolation of $\kappa_c(\beta)$ from ETMC results.}
\label{fig:action}
\end{figure}

Wilson-type fermions suffer from unphysical phases in their bare parameter phase space. For standard Wilson fermions the Aoki phase at strong coupling is well known~\cite{aoki:1984mm,aoki:1986mm}. For the enlarged $(\kappa,\beta,\mu)$ space of twisted mass fermions an additional phase transition plane has been found that incorporates the line of critical hopping parameters $\kappa_c(\beta,\mu=0)$~\cite{Sharpe:2004ps,Farchioni:2004us,Farchioni:2005ec}.
      
It is of course very important to make sure that the finite temperature simulations are not affected by these unphysical phases. In Fig.~\ref{fig:action}, left, our findings are summarised~\cite{Ilgenfritz:2009ns}. 
The thermal transition can be found as a surface that winds around the line of critical hopping parameters. 
This behaviour can be understood as a remnant of the continuum symmetry under twist rotations~\cite{Creutz:2007fe}. In conclusion, it is possible to perform  physical simulations towards the continuum limit that are not affected by unphysical bulk behaviour.

Our simulation strategy is to perform scans in the lattice coupling $\beta$ while keeping maximal twist in order to be $\mathcal{O}(a)$ improved. The knowledge of $\kappa_c(\beta)$ as well as the lattice spacing and pion masses can be obtained from ETMC data~\cite{Baron:2009wt,personal}. 
The interpolation of $\kappa_c(\beta)$ is shown in Fig.~\ref{fig:action}, right. It is also possible to keep a fixed pion mass by adjusting $\mu(\beta,m_\pi)$. 

\section{Thermal signals}\label{sec:thermal}

\begin{figure}
\includegraphics[width=.35\textwidth,angle=-90]{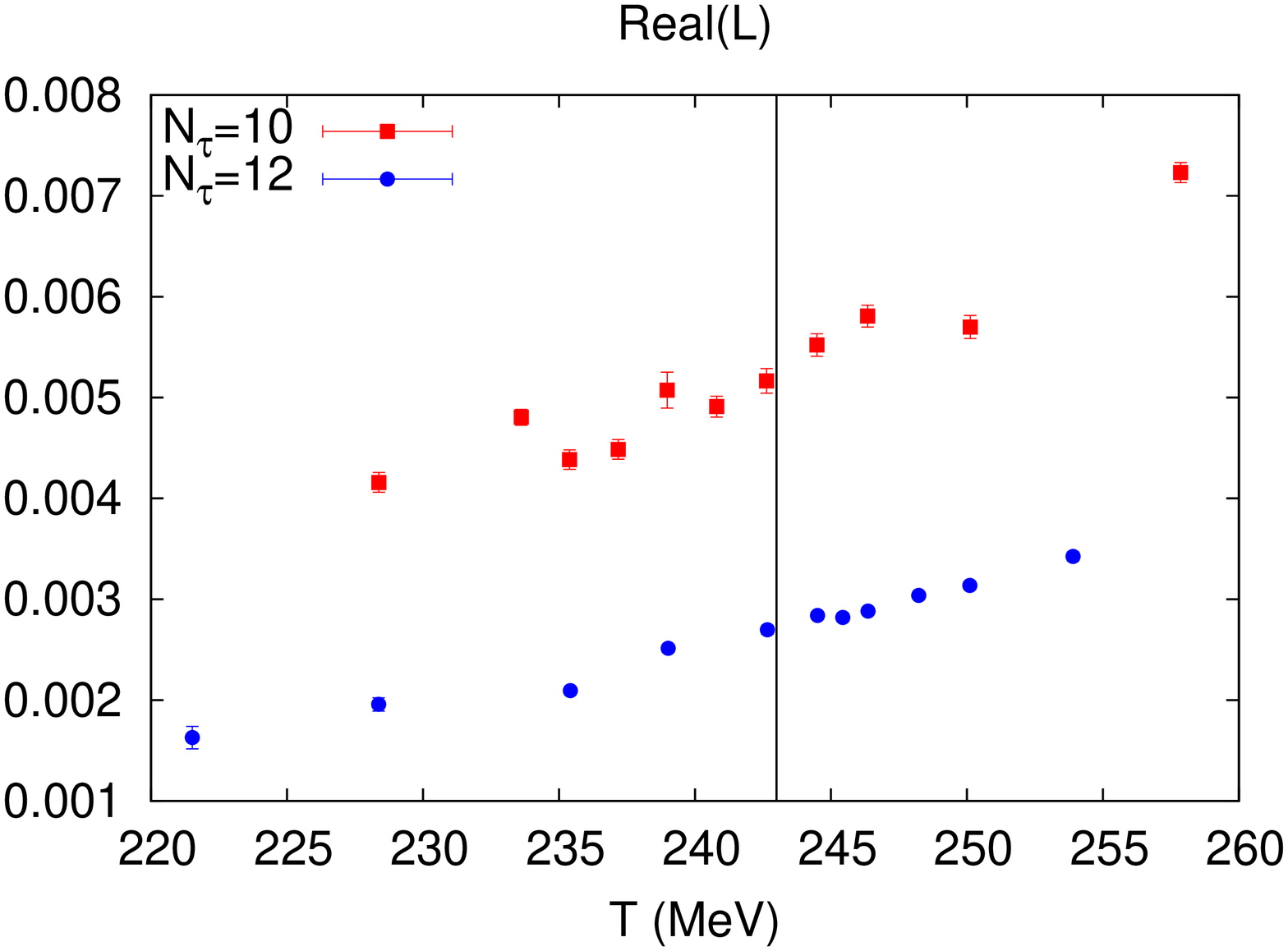}
\hfill
\includegraphics[width=.35\textwidth,angle=-90]{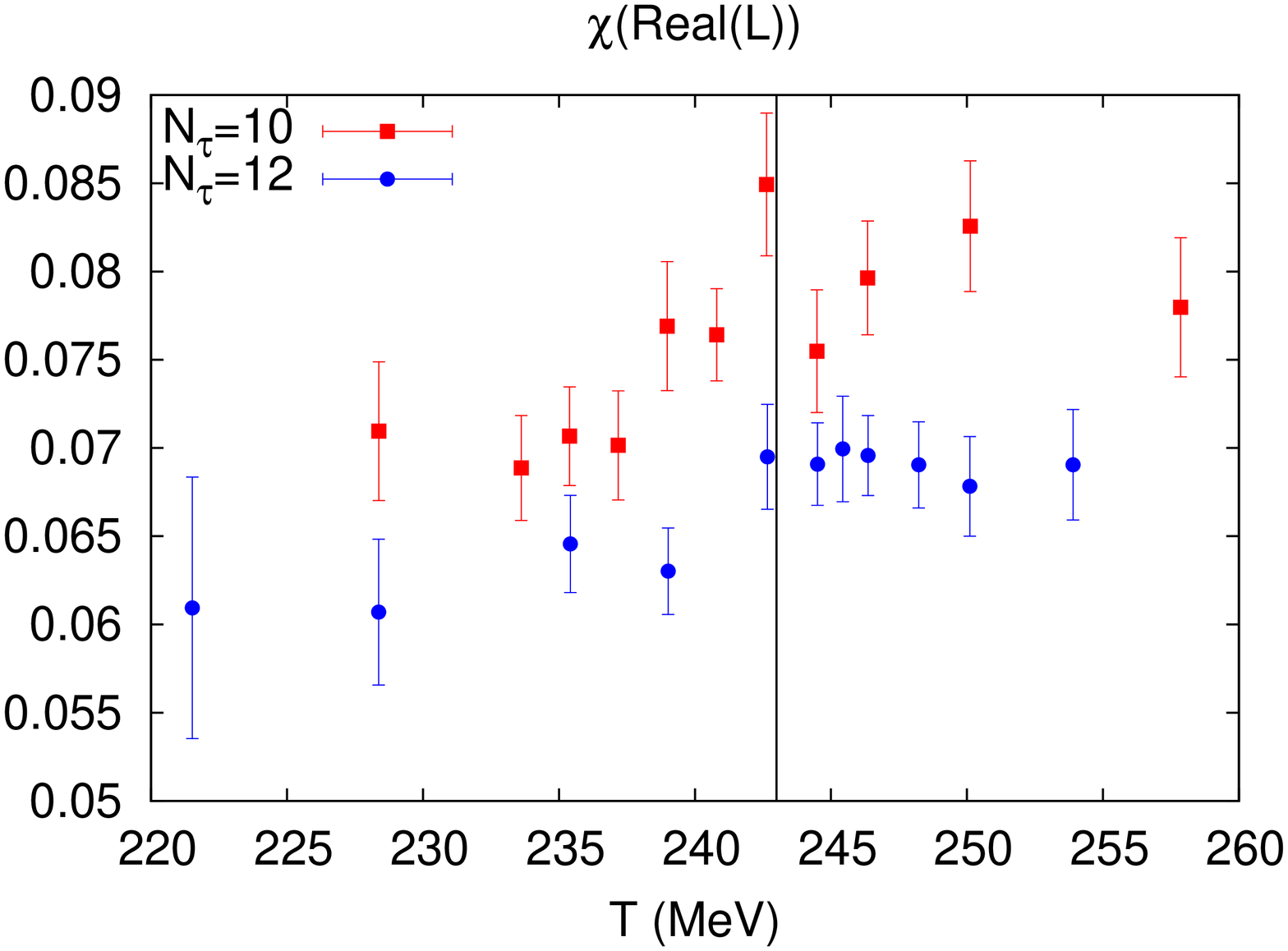}
\caption{Real part of the Polyakov loop (left) and its susceptibility (right) from the two available lattice spacings at $m_\pi\approx380$\,MeV. The vertical line indicates the value of $T_c$.}
\label{fig:polyakov}
\end{figure}

\begin{figure}
\centering
\includegraphics[width=.34\textwidth,angle=-90]{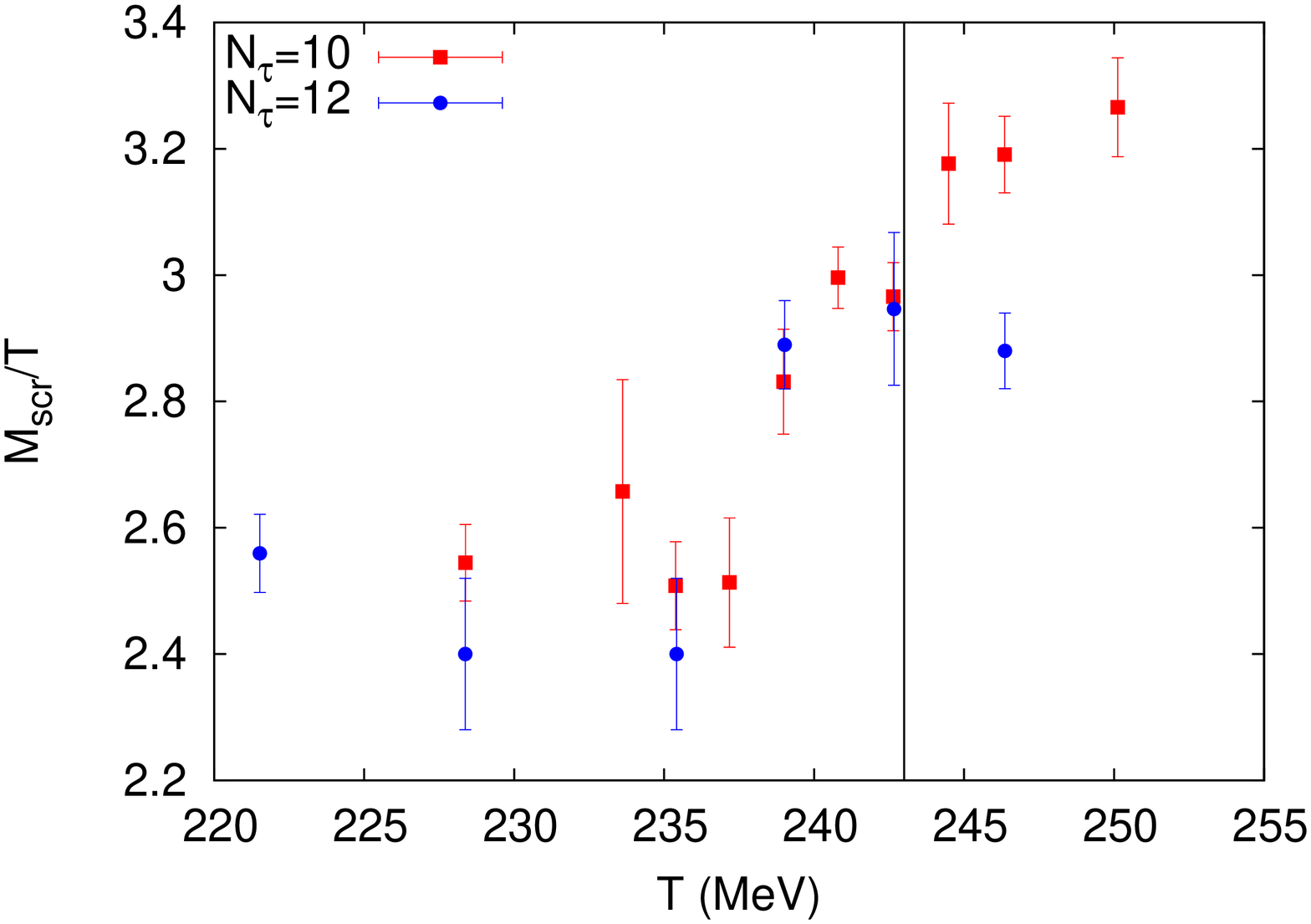}
\hfill
\includegraphics[width=.34\textwidth,angle=-90]{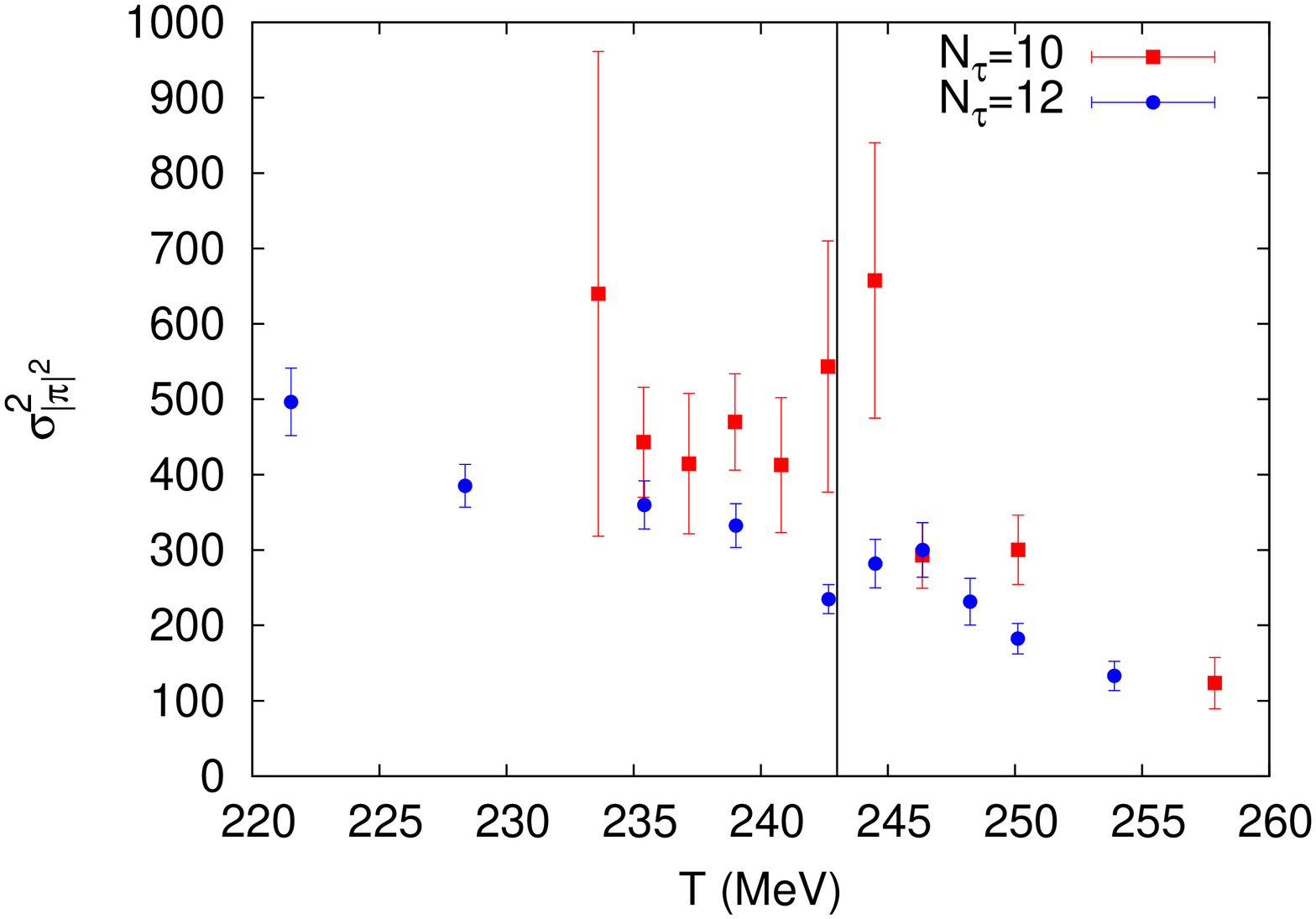}
\caption{Left: Flavour non-singlet pseudo-scalar screening mass from the two available lattice spacings at $m_\pi\approx380$\,MeV.
Right: Variance of pion norm at $m_\pi\approx380$\,MeV. The vertical line marks $T_c$ from the Polyakov loop susceptibility.}
\label{fig:mscrandpi}
\end{figure}

Since the expected behaviour is that of a continuous crossover, it is very important to study the temperature dependence of several observables. In our study, we look at the real part of the Polyakov loop, the chiral condensate, the pion norm (pseudo-scalar susceptibility), the plaquette and screening masses.

In this contribution we present plots for the Polyakov loop, Fig.~\ref{fig:polyakov}, the pseudo-scalar flavour non-singlet screening mass and the variance of the pion norm, Fig.~\ref{fig:mscrandpi}, at $m_\pi\approx380$\,MeV.
The peak position of the susceptibility of the Polyakov loop is used to define $T_c$. For $m_\pi\approx 380$\,MeV, we obtain $T_c(N_\tau=10)=243(10)$\,MeV and $T_c(N_\tau=12)=243(8)$\,MeV from the respective maximal values.

These crossover temperatures are consistent with the pseudo-scalar flavour non-singlet screening mass, presented in Fig.~\ref{fig:mscrandpi}, left. This screening mass shows a qualitative change in behaviour at the onset of the transition region, i.\,e. just left to the corresponding value of $T_c$. A combined analysis of this screening mass and the corresponding scalar one will give information on the splitting introduced by the $U_A(1)$-anomaly. 
If the anomaly is still sufficiently strong near $T_c$, the transition in the chiral limit is expected to be in the $O(4)$ universality class, although a first order scenario is not excluded~\cite{Pisarski:1983ms,ParisenToldin:2003hq}.
 The size of this splitting is thus an indicator for the order of the $N_f=2$ transition in the chiral limit.

The variance of the pion norm normalised to the spatial volume
\begin{equation}
\sigma^2_{\vert\pi\vert^2} = N_\sigma^3 \left( \left< \left(\vert\pi\vert^2\right)^2\right> - \left<\vert\pi\vert^2 \vphantom{\left< \left(\vert\pi\vert^2\right)^2\right> }  \right>^2\right)
\end{equation}
is shown in Fig.~\ref{fig:mscrandpi}, right. For this observable a peak can be found at approximately the same position at which the Polyakov susceptibility peaks.

\begin{figure}
\includegraphics[width=.35\textwidth,angle=-90]{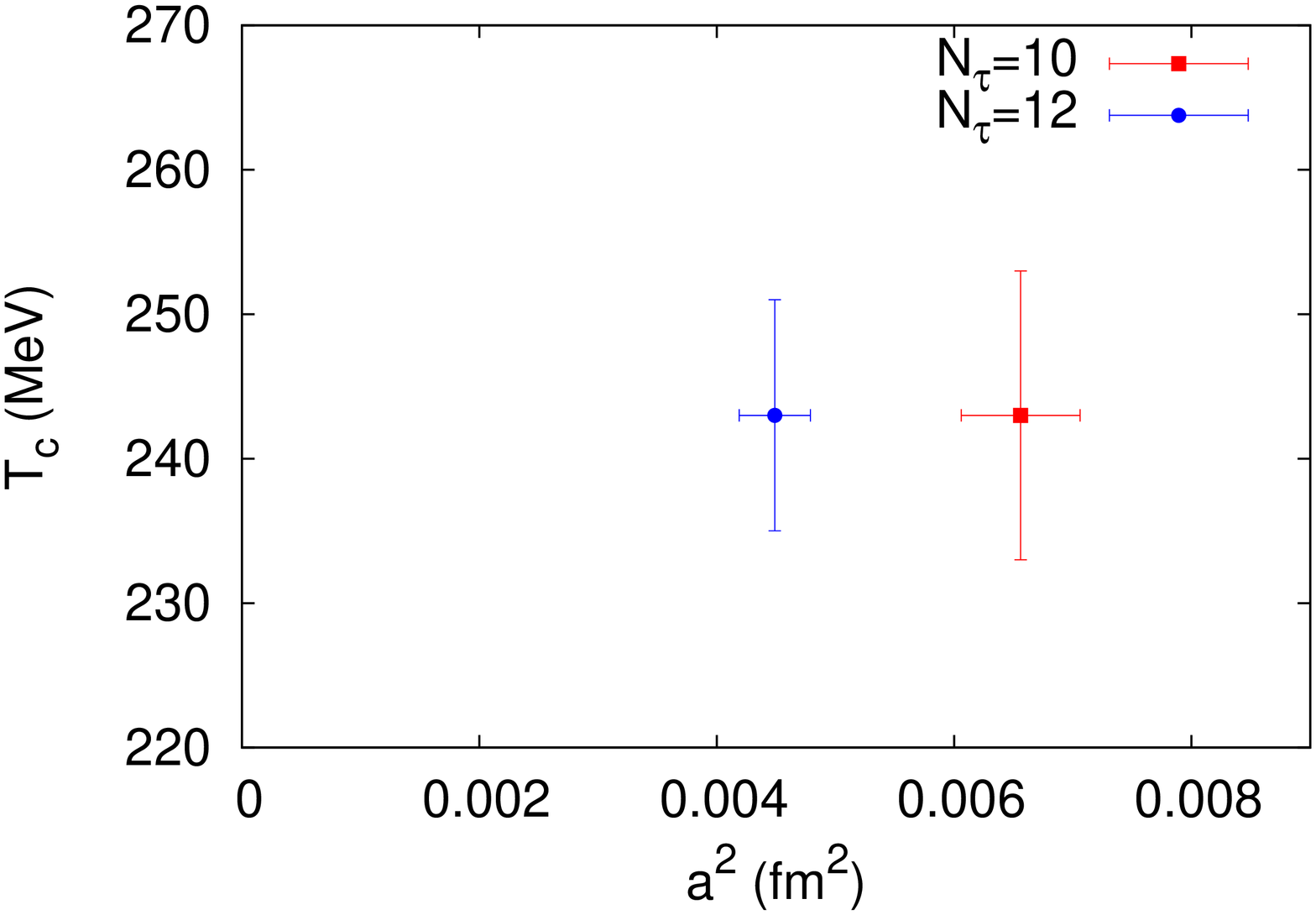}
\hfill
\includegraphics[width=.35\textwidth,angle=-90]{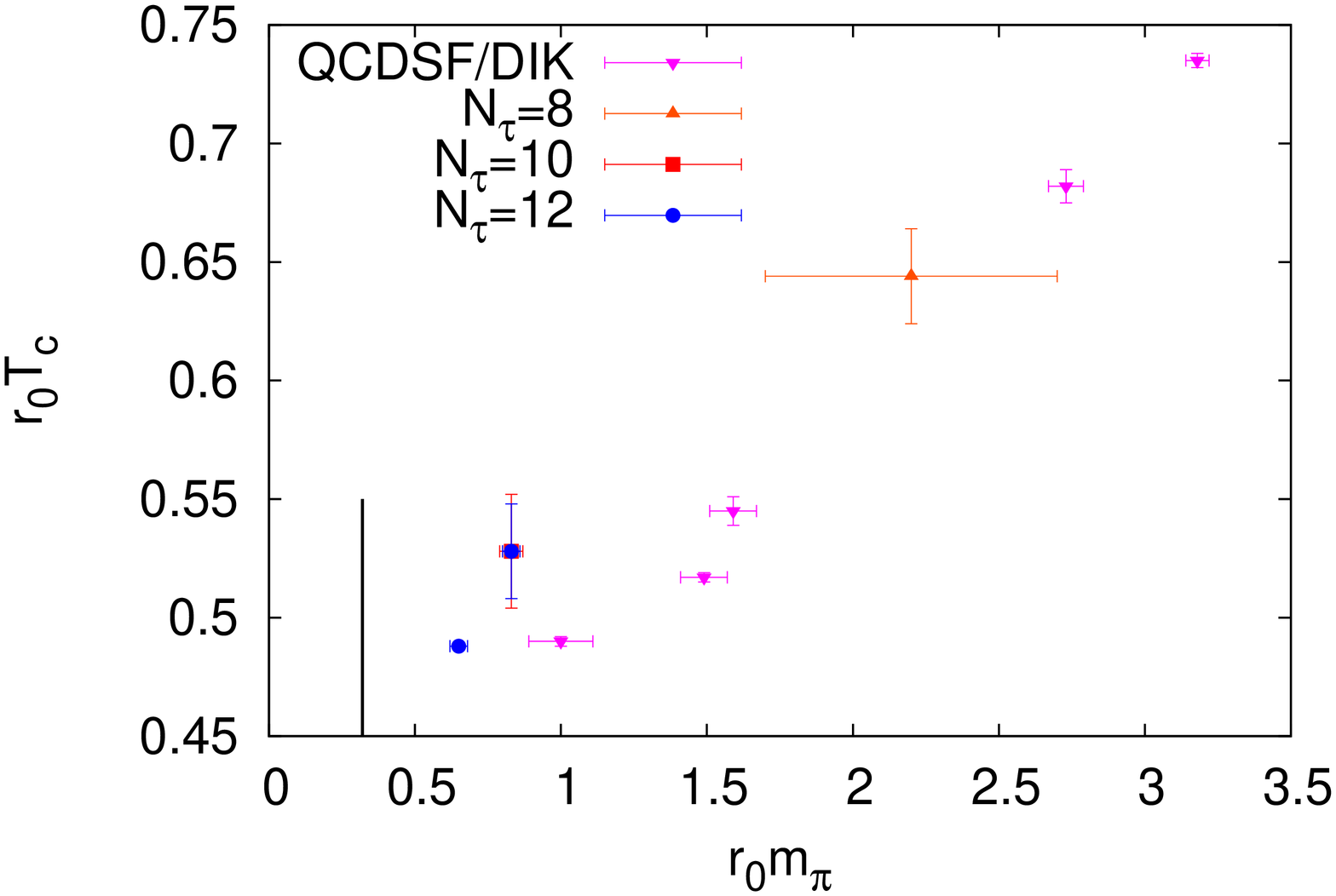}
\caption{
Left: Lattice spacing dependence for $T_c$ as detemined by the Polyakov loop for $m_\pi\approx380$\,MeV.
Right: Values of $T_c(m_\pi)$ from the Polyakov loop susceptibility for different lattice spacings. The values of QCDSF/DIK from~\cite{Bornyakov:2009qh} are shown for comparison. 
}
\label{fig:tc}
\end{figure}

In Fig.~\ref{fig:tc}, left, $T_c$ for the two lattice spacings at $m_\pi\approx380$\,MeV is plotted. The $N_\tau=12$ result can possibly be further refined and at least a third lattice spacing is needed for a final statement. However, the cutoff dependence appears to be rather small.

In addtion to the points for $m_\pi\approx 380$\,MeV we currently simulate at $N_\tau=12$ with $m_\pi\approx300$\,MeV and 460\,MeV. For $m_\pi\approx300$\,MeV, a first value of $T_c\approx225(11)$\,MeV is available. 
Fig.~\ref{fig:tc}, right, shows our results compared to those of~\cite{Bornyakov:2009qh}. It seems that our points tend to be at somewhat higher temperatures. A small shift in our results due to the scale set by the new ETMC data might occur, but we expect this to be at the percent level.

\section{Conclusions and outlook}\label{sec:concl}
We have presented results of our study of the thermal transition of QCD applying maximally twisted mass fermions. 
Relying on the ETMC data at zero temperature, we can investigate a mass range $m_\pi \gtrsim 300$\,MeV. 
Our current simulations will soon provide enough input to perform a chiral and continuum extrapolation of the critical temperature. So far we have concentrated on the Polyakov loop to define this temperature. 
Screening masses supplement our set of observables. Their splittings can give insights into the $N_f=2$ chiral transition.

\acknowledgments
O.P. and L.Z. acknowledge support by the Deutsche Forschungsgemeinschaft, grant PH 158/3-1.
F.B. and M. M.-P. acknowledge financial support by
DFG (GK 1504 and SFB/TR 9).
We thank for the generous support with supercomputing power
by the HLRN Berlin and Hannover. 
Some of the computations have been performed on the apeNEXT in Rome. We thank the apeNEXT staff for continual support.

\end{document}